\documentclass[prl,twocolumn,runinaddress,showpacs]{revtex4}
\usepackage{amsmath,amsthm,amssymb,amscd,dcolumn,float}
\usepackage{psfrag,graphicx}
 \def\R{\mathbf R}
\def\x{\mathbf x} 
\let\d\partial
 \def\z{\mathbf z}   
 \let\a\alpha \let\be\beta  
\let\t\tau 
\def\Hb{\,\overline{\!H}}

\def\P{\mathcal P}
 \def\ii{\mathrm i}
\def\U{\mathcal U}
\def\N{\mathcal N}

\def\M{\mathcal M}

\def\D{\mathcal D}
\def\operator#1{\expandafter\def\csname#1\endcsname{\operatorname{#1}}}
\operator{sech} \operator{csch}

\theoremstyle{definition}

\theoremstyle{remark}

\def\({{\rm(}} \def\){{\rm)}}
\begin{document}
\title{Exact solutions of an elliptic Calogero--Sutherland
model}
\author{D. G\'omez-Ullate}
\email{dgu@eucmos.sim.ucm.es}
\author{%
A. Gonz\'alez-L\'opez}
\email{artemio@eucmos.sim.ucm.es}
\author{M. A. Rodr\'{\i}guez}
\email{rodrigue@eucmos.sim.ucm.es}
\affiliation{Departamento de F\'{\i}sica Te\'orica II,
Facultad de Ciencias F\'{\i}sicas,
Universidad Complutense,
28040 Madrid, Spain}
\date{June 5, 2000}
\begin{abstract}
A model describing $N$ particles on a line interacting pairwise
via an elliptic function potential in the presence of an external field is
partially solved in the quantum case in a totally algebraic way.
As an example, the ground state and the lowest excitations are
calculated explicitly for $N=2$.
\end{abstract}
\maketitle
It is well known that the class of exactly solvable problems does
not include most physical problems. The development of computer
science in the last decades has made possible the use of numerical
methods to approximate exact solutions in a wide variety of
situations. Yet, the study of exactly solvable models still
deserves attention, not only because the knowledge of exact
solutions can be used to test approximate methods, but also in its
own right, due to the simplicity and mathematical beauty of the
models, and the wide range of connections with other fields of
physical and mathematical research.

This is illustrated by the renewed interest in the
Calogero--Sutherland (CS) models of interacting particles in
one dimension, which have been recently applied to many different
fields such us quantum spin chains with long range interaction
\cite{spin}, random matrix theory \cite{matrix}, fractional statistics
and anyons \cite{anyons}, Yang-Mills theories \cite{YM}, quantum Hall
liquids \cite{QH}, soliton theory \cite{Pol}, vicinal surfaces in
crystals \cite{Las}, and black holes \cite{GT}.

The first example of a non-trivial integrable quantum many-body
problem was found by Calogero \cite{Cal1}, and consists of a system of
identical nonrelativistic particles interacting pairwise through an
inverse-square potential $v(r) = r^{-2}$, so that
\begin{equation}\label{An}
  H_N = -\sum_{k=1}^N \d^2_{x_k} + g \sum_{\substack{j,k=1\\j \ne
  k}}^N v(x_j-x_k)\,.
\end{equation}
By \emph{integrable} we mean here that a complete commuting set of
constants of motion can be explicitly constructed.
Soon afterwards, Sutherland \cite{Suth} established the integrability of
the model \eqref{An} with an inverse sine-square interaction $v(r) =
\sin^{-2}(r)$.

The most general interaction potential $v$ for which the Hamiltonian
\eqref{An} is known to be integrable is the Weierstrass $\P$ function,
which includes the rational and trigonometric cases as special limits.
The integrability of this potential was proved in the classical case
by Calogero and Perelomov by means of a Lax pair representation
\cite{CP}, and its explicit integration was performed by Krichever,
\cite{Krich}. Olshanetsky and Perelomov \cite{OP} later showed that
all these models have an underlying algebraic structure based on root
systems of $A_n$ algebras, and that integrable models associated to
other root systems also exist. In the models treated in Ref.~\cite{OP}
the integrals of motion are related to the radial parts of the
Laplace--Beltrami operator on a symmetric space associated to the
given root system. These integrable models are obtained from the
projection of free motion on a higher-dimensional manifold.

However, integrable Hamiltonians are not necessarily \emph{solvable},
i.e., we might not be able to find explicitly their spectrum and
eigenfunctions. The models with inverse-square and inverse sine-square
interaction are known to be solvable, and much literature has been
devoted to the study of their eigenfunctions \cite{Jack}, but the more
general model with the Weierstrass $\P$ potential is considerably more
difficult. In fact, very few explicit solutions are known in this case
\cite{Per}, and only for a low number of particles.

The purpose of this Letter is to present a model of $N$ particles on a
line with elliptic pairwise interaction in an external field for which
a finite number of eigenvalues and eigenfunctions can be computed
algebraically. We shall only sketch here the main ideas behind the
proof of this result, referring the reader to our previous work
\cite{JPA} for a more complete description of the method used.

Consider the $N$-body quantum Hamiltonian
\begin{equation}\label{ham2}
H_N =-\sum_{k=1}^N \d_{x_k}^2+ V_N (\x)
\end{equation}
with potential
\begin{align}
    \label{potential}
V_N(\x)&= c_m \sum_{k=1}^N \P(x_k + \ii \be) + 4b (b-1)\sum_{k=1}^N
\P(2x_k)\notag\\
&\enspace {} + a(a-1) \sum_{\substack{j,k=1\\j \ne k}}^N
[ \P(x_j + x_k) + \P(x_j-x_k)]\,,
\end{align}
where
$a$ and $b$ are positive real parameters, $m$ is a non-negative
integer,
\begin{multline}
c_m= 2\big[2b+m+a(N-1)\big]\\\times\big[2b+2m+2a(N-1)+1\big]>0\,,
\end{multline}
and $\P(z)\equiv\P(z|g_2,g_3)$ denotes the Weierstrass $\P$
function with invariants $g_2,g_3\in\R$. If $g_2$ and $g_3$ satisfy
the inequality $g_2^3>27 g_3^2$, then $\P(z)$ has two fundamental
periods $2\alpha$ and $2 \ii \be$ which are real and purely imaginary,
respectively \cite{WW}. In this case $\P(x+\ii\be)$ is real and regular
(analytic) for all real values of $x$, with real period $2\a$. On the
other hand, $\P(x)$ is real for real $x$ and diverges as
$(x-2n\a)^{-2}$ when $x$ tends to an integer multiple $2n\a$ of the real
period $2\a$. Thus, the configuration space for the Hamiltonian
\eqref{ham2}--\eqref{potential} can be taken as the bounded region of
$\R^N$
\begin{equation}\label{region}
 0<x_N<x_{N-1}<\dots<x_1<
\alpha\;.
\end{equation}
Since the potential \eqref{potential} is confining in this region, the
spectrum of $H_N$ is purely discrete, and the boundary condition satisfied 
by its eigenfunctions $\psi_k({\bf x})$ is their vanishing on the boundary of
\eqref{region}.

The potential \eqref{potential} with $c_m=0$ is of $C_N$ type \cite{OP}
$$
4b(b-1)\sum_{k=1}^N v(2x_k)+a(a-1)\sum_{\substack{j,k=1\\j \ne 
k}}^N\big[v(x_j-x_k)+v(x_j+x_k)\big]\,,
$$
with interaction potential $v(r)=\P(r)$. The term proportional to
$c_m$ in \eqref{potential} can be viewed as the contribution of an
external field with potential $\P(r+\ii\be)$.

We shall now show that, when the parameter $m$ is a non-negative
integer, one can algebraically compute a finite number (depending on
$m$, see \eqref{dim} below) of eigenvalues and eigenfunctions of the
Hamiltonian \eqref{ham2}--\eqref{potential}. These {\em algebraic
eigenfunctions} have the form
\begin{equation}
    \psi_k(\x)=\mu(\x)\,\chi_k({\bf z})\,,
    \label{eigenf}
\end{equation}
where
\begin{equation}\label{mu}
\mu (\x) = \prod_{j<k}\left[ \P(x_j+\ii\beta)-\P(x_k+\ii\beta)
\right]^a \,\prod_k \left[ \P'(x_k+\ii\beta) \right]^b\,,
\end{equation}
and $\chi_k({\bf z})$ is a suitable completely symmetric polynomial of
degree at most $Nm$ in the variables
\begin{equation}\label{cofv}
z_j= \P(x_j+\ii\beta),\quad\,j=1,\dots, N\,.
\end{equation}
The exact solutions of the trigonometric and
rational $C_N$ (in fact, $BC_N$) models also assume the form
\eqref{eigenf}, but in this case $\mu$ can be factorized over the
system of positive roots, i.e., it has the form
\begin{equation}
    \label{factor}
\mu = \prod_{j<k} \big[ f(x_j-x_k) f(x_j+x_k) \big]^a \prod_k
\big[f(2x_k)\big]^b\,,
\end{equation}
where $f(x) = x$ in the rational case and $f(x) = \sin x$ in the
trigonometric case. Moreover, in both cases $\mu$ coincides with the
ground-state wave function of the system. By contrast, the function
$\mu$ in \eqref{mu} cannot be factorized over the system of positive
roots for any function $f(x)$, and is not the ground-state
wave function of the Hamiltonian \eqref{ham2}--\eqref{potential}. As a
matter of fact, it was shown in \cite{Cal2} that the most general
potential allowing for the factorization
\eqref{eigenf}--\eqref{factor} does not include the elliptic case.
This is one of the reasons why it has been so difficult to obtain
explicit solutions of the elliptic CS models.

When the parameters $a$ and $b$ are positive, the functions
\eqref{eigenf} are regular in the region \eqref{region}, and they
automatically vanish on its boundary on account of the identities
$\P'(\ii\be)=\P'(\a+\ii\be)=0$ (see Ref.~\cite{WW}). Thus, to show
that $\psi_k$ in Eq.~\eqref{eigenf} is an eigenfunction of $H_N$ we
only have to check that it satisfies the Schr\"odinger equation
$(H_N-E_k)\psi_k=0$ in the open region \eqref{region}. Equivalently,
$\chi_k$ must be a solution of the equation $(\Hb_N-E_k)\chi_k=0$,
where the {\em gauge Hamiltonian} $\Hb_N$ is defined by
\begin{equation}\label{gaugeham}
\Hb_N = \mu^{-1}\, H_N\,\mu\,.
\end{equation}
Note that, by the standard properties of the Weierstrass function
\cite{WW}, $\mu$ does not vanish in the region \eqref{region}.

It can be shown that, provided that $m$ is a non-negative
integer, $\Hb_N$ preserves the finite-dimensional
polynomial space
\begin{equation}\label{module}
\M_m = \text{span} \left\{ \t_1^{l_1}\t_2^{l_2} \cdots
\t_N^{l_N}:\sum_{i=1}^N l_i \leq m\right\},
\end{equation}
where
\begin{equation}
\t_k= \sum_{i_1<i_2< \dots <i_k} z_{i_1}z_{i_2}\cdots z_{i_k}\,,
\qquad 1\le k\le N\,,
\end{equation}
are the elementary symmetric functions of the variables $z_k$. This is
due to the fact that, when $\Hb_N$ is written in terms of the
symmetric variables $\t_1,\dots,\t_N$, it can be expressed as a
quadratic combination of the generators of $sl(N+1)$ in the
representation
\begin{align}
    \D_k&=\d_{\t_k}\,,\quad
    \N_{jk}=\t_j\,\d_{\t_k}\,,\quad
    \U_k=\t_k\left(m-\sum_{i=1}^N \t_i\,\d_{\t_i}\right);\notag\\
\label{slN}
    j,k&=1,2,\dots,N\,.
\end{align}
Since these generators obviously preserve the subspace
\eqref{module}, so does the gauge Hamiltonian $\Hb_N$. It follows that
$\Hb_N$ has (at most)
\begin{equation}\label{dim}
\dim \M_m = \binom{m+N}m
\end{equation}
eigenfunctions $\chi_k$ lying in $\M_m$, which can be algebraically
computed, along with their corresponding eigenvalues, simply by
diagonalizing the finite-dimensional matrix of $\Hb_N|_{\M_m}$. The
elements of $\M_m$, being polynomials in the symmetric variables
$\t_k$ of degree at most $m$, are symmetric polynomials in $\z$ of
degree not greater than $Nm$. Thus the original Hamiltonian
\eqref{ham2}--\eqref{potential} possesses (at most) $\dim\M_m$
algebraically computable eigenfunctions of the form
\eqref{eigenf}--\eqref{mu}, with $\chi_k(\z)$ a symmetric polynomial
of degree at most $Nm$, as claimed. Note, however, that there are
other eigenfunctions of \eqref{ham2} which do not belong to this
algebraic sector. An interesting open problem is to analyze the
position in the spectrum of the algebraic eigenvalues and their
degeneracy. The algebra $sl(N+1)$, which plays a fundamental role in
the partial integrability of the Hamiltonian
\eqref{ham2}--\eqref{potential}, is sometimes called a \emph{hidden
symmetry algebra} \cite{QESrefs}, since in this case the Hamiltonian
need not be a Casimir element.

For simplicity (indeed, no conceptual
difficulties arise for higher values of $N$ and $m$), let us consider
the problem for $N=2$ and $m=2$, for which the potential reads
\begin{multline}
    V_2(x_1,x_2) = c_2 \sum_{k=1}^2 \P(x_k + \ii \beta) + 4b
    (b-1)\sum_{k=1}^{2} \P(2x_k) \\
    {}+2 a(a-1) [ \P(x_1 + x_2) + \P(x_1-x_2)]\,.
    \label{V2}
\end{multline}
Note that this is intrinsically a two-body problem, since the
potential \eqref{V2} is not translation invariant.
The number of algebraic eigenstates is  at most
$\dim \M_2 = 6$, and the matrix of the restriction $\Hb_2 |_{\M_2}$ with
respect to the canonical basis $\{1,\t_1,\t_2,\t_1^2,\t_1\t_2,\t_2^2\}$ of
$\M_2$ is given by
\begin{widetext}
\begin{small}
\begin{equation}
    \label{matrix}
\begin{pmatrix}
 0 & g_2(2a+2b+1) & -2 a g_3 & 4 g_3 & 0 & 0 \\
16a+24b+20 & 0 & g_2(b+1/2) & 4g_2(a+b+1) & 2 g_3(1-a) & 0 \\
 0 & 8a+24b+12 & 0 & 0 & g_2(2a+2b+5) & -4g_3(a+1) \\
0 & 8a+12b+14 & 0 & 0 & g_2(b+1/2) & 2g_3 \\
 0 & 0 & 8a+12b+14 & 16(a+3b+3) & 0 & g_2 (2b+3) \\
0 & 0 & 0 & 0 & 8a+24b+28 & 0
\end{pmatrix}.
\end{equation}
\end{small}
\end{widetext}
The eigenvalues of this matrix are the algebraic energies of the 
physical Hamiltonian $H_2$, and its eigenvectors give the components 
of the corresponding functions $\chi_k$ in Eq.~\eqref{eigenf} with 
respect to the canonical basis of $\M_2$.

\begin{figure}
\centering
\includegraphics[width=7.5 cm]{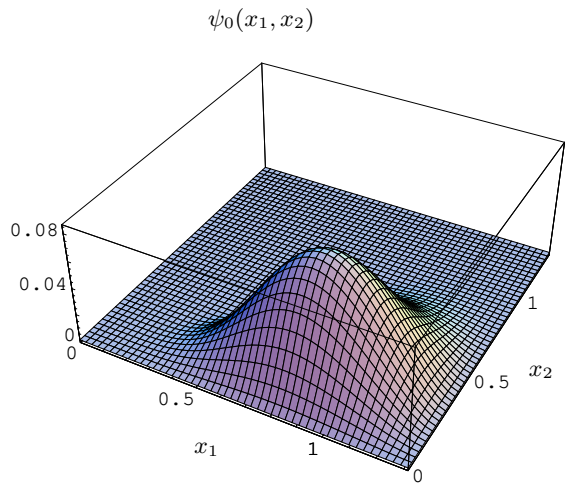}
\caption{Ground-state wave function of the potential \eqref{V2}.}
\label{ground}
\end{figure}

\begin{figure}
  \centering
\includegraphics[width=7.5 cm]{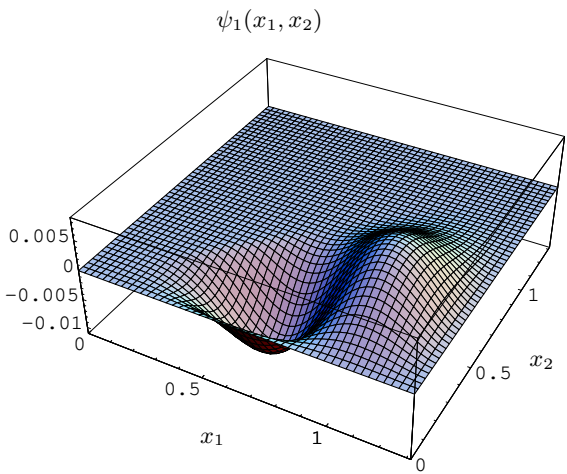}
\caption{First excited state wave function of the potential \eqref{V2}.}
\label{first}
\end{figure}

\begin{figure}
  \centering
\includegraphics[width=7.5 cm]{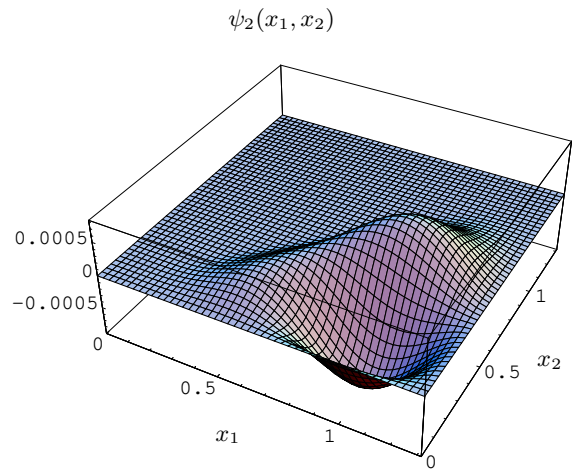}
\caption{Second excited state wave function of the potential \eqref{V2}.}
\label{second}
\end{figure}

Take, for instance, the following values of the coupling constants and the
invariants of $\P(x)$:
\begin{equation}
    a=2\,,\qquad
    b=3/2\,,
    \qquad
    g_2=3\,,
    \qquad
    g_3=2/3\,,
    \label{params}
\end{equation}
for which the half-periods are
$$
\alpha = 1.31523,\qquad \ii\,\beta
=1.61809\,\ii\,.
$$
The matrix \eqref{matrix} has six real eigenvalues, so that the
Hamiltonian \eqref{ham2}--\eqref{V2} possesses six algebraic energies, which
have been listed in Table \ref{table}.
\begin{table}[b]
\begin{tabular}{cdd}\toprule
 & \multicolumn{1}{c}{\quad Exact} & \multicolumn{1}{c}{Numerical}\\
 \colrule
$E_0$&  \quad -86.5484\qquad &\qquad -86.40\quad  \\
\colrule
$E_1$ & -43.2786 & -42.96\\
\colrule
$E_2$ & -10.0288 & -9.64  \\
\colrule
$E_3$ & 12.5045 & 13.11\\
\colrule
$E_4$ & 46.2657 & 46.97\\
\colrule
$E_5$ & 81.0857 & 81.34  \\
\botrule
\end{tabular}
\caption{Algebraic eigenvalues $E_i$ of the Hamiltonian 
\eqref{ham2}--\eqref{V2} and their numerical approximations.}
\label{table}
\end{table}
\noindent In this table the (exact) algebraic eigenvalues of $H_2$
have been compared with a numerical approximation of the six lowest
energy levels obtained by solving the Schr\"odinger equation in the
triangle $0<x_2<x_1<\alpha$ using a finite element method. The
approximate agreement between the two columns of Table \ref{table}
shows that the algebraic eigenvalues are in this case the lowest
energy states of the system, although there is no guarantee that this
should still be true in the general case. The polynomials $\chi_k$
corresponding to the six algebraic eigenfunctions of the potential
\eqref{V2}--\eqref{params} are given in Table \ref{eigtable}. In
Figs.~\ref{ground}--\ref{second} we present a plot of the ground state
and the first two excited wave functions of the system.
\begin{table}[h]
\begin{tabular}{crrrrr}
    \toprule 
    &\multicolumn{1}{c}{$\tau_1$} &
    \multicolumn{1}{c}{$\tau_2$} &
    \multicolumn{1}{c}{$\tau_1^2$} &
    \multicolumn{1}{c}{$\tau_1\tau_2$} &
    \multicolumn{1}{c}{$\tau_2^2$\vrule height 9pt width 0pt}\\
    \colrule
    $\chi_0\quad$ & $-3.0585$ & $7.1643$ & $2.2349$ & 
    $-9.778$ & $9.0382$\\
    \colrule
    $\chi_1\quad$ & $-3.3008$ & $-10.917$ 
    & $2.5612$ & $13.462$ & $-24.885$\\
    \colrule
    $\chi_2\quad$ & $0.40412$ & $5.1067$ & $-2.2912$ 
    & $-0.77221$ & $6.1599$\\
    \colrule
    $\chi_3\quad$ & $-3.9273$ & $-37.989$
    & $2.0457$ & $14.734$ & 
    $94.264$\\
    \colrule
    $\chi_4\quad$ & $1.6422$ & $-2.7456$ & $-0.17611$ 
    & $-10.472$ & $-18.108$\\
    \colrule
    $\chi_5\quad$ & $3.8942$ & $8.3079$ 
    & $3.6675$ & $15.1$ & $14.898$\\
    \botrule
\end{tabular}
\caption{Polynomials $\chi_k$ corresponding to the algebraic
eigenfunctions $\psi_k=\mu\,\chi_k$ of the Hamiltonian 
\eqref{ham2}--\eqref{V2}. In all cases, the coefficient of $1$ has 
been normalized to unity.}
\label{eigtable}
\end{table}

The fact that $\psi_0=\mu\chi_0$ is the ground state of the system is 
immediately apparent if we note that $\chi_0$ can be expressed in 
terms of the variables $z_k$ as
\begin{align}
    \chi_0 &= 2.2349 (z_2-0.82835) (z_2-0.54017)\notag\\
    &\quad{}-9.778\,z_1\,(z_2-0.79769) (z_2-0.39212)\notag\\ 
    &\quad{}+ 9.0382\,z_1^2(z_2-0.75385) (z_2-0.32801).
\end{align}
For the values of the invariants $g_2$ and $g_3$ given in
\eqref{params}, we have \cite{WW}
$$
e_3=-0.72011\le z_k=\P(x_k+\ii\be)\le
e_2=-0.240851\,.
$$
We thus see from the previous expression that $\chi_0$ is positive
everywhere. Since, by \eqref{factor}, $\mu$ has no zeros in the open triangle
$0<x_2<x_1<\a$, it follows that $\psi_0$ does not vanish in this triangle.
\psfrag{p}[Bc][Bc][1][0]{$\psi_0(x_1,x_2)$}
\psfrag{x}{$x_1$}
\psfrag{y}{$x_2$}
\psfrag{p}[Bc][Bc][1][0]{$\psi_1(x_1,x_2)$}
\psfrag{p}[Bc][Bc][1][0]{$\psi_2(x_1,x_2)$}

In conclusion, a finite number of eigenvalues and eigenfunctions of a
quantum Hamiltonian describing $N$ particles on a line with elliptic
interaction in the presence of an external field have been explicitly
calculated by an algebraic method independent of the usual approach
based on root systems.
\begin{acknowledgments}
  The authors gratefully acknowledge the partial financial support
  of the {\em Direcci\'on General de Ense\~nanza Superior} under
  grant DGES PB98--0821. They would also like to thank A. M. Perelomov 
  and O. Ragnisco for several useful conversations.
\end{acknowledgments}

\end{document}